\documentclass[runningheads]{svmult}

\usepackage{graphicx}  

\begin{document}
\title*{The Atmospheric Cherenkov Imaging Technique for Very 
High Energy Gamma-ray Astronomy}
\titlerunning{ACIT}
\author{Trevor C. Weekes}
\authorrunning{Trevor C. Weekes}
\institute{Whipple Observatory, Harvard-Smithsonian Center for Astrophysics,
            P.O. Box 6369, Amado, Arizona 85645-0097, U.S.A.; 
e-mail:tweekes@cfa.harvard.edu}

\maketitle              

\begin{abstract}

The Atmospheric Cherenkov Imaging Technique has opened up the gamma-ray 
spectrum
from 100 GeV to 50 TeV to astrophysical exploration. The development of the 
technique is described as are the basic principles underlying its use. 
The current
generation of arrays of telescopes is briefly described and the early results 
are
summarized.
\footnote{Written Version of Lectures given at the International 
Heraeus Summer School
on ''Physics with Cosmic Accelerators'', Bad Honnef, Germany, July 5 - 16, 2004
(to be published by Springer-Verlag in their Lecture Notes Series).} 
\end{abstract}

\section{Introduction}

One of the last frontiers of the gamma-ray sky is that characterized by 
the distribution of
TeV photons. These photons can be detected relatively easily with ground-based
detectors (constituting a TeV ''window'' in the atmosphere) and thus the 
detection of sources did not have
to await the availability of space platforms. In practice although the 
technology 
was available 
at an early date, it required the impetus of gamma-ray space astronomy to 
justify a major
effort in a new discipline. Since it concerns the highest energy photons 
with which 
it is yet 
feasible to map the sky, it is of particular interest to high energy 
astrophysics. 
Any source of TeV photons must be associated with a cosmic
particle accelerator and of inherent interest to high energy particle 
physicists as well as
students of the cosmic radiation. 

To date almost all the observational results in the energy interval
100 GeV - 100 TeV have come from observations using the so-called 
''Atmospheric Cherenkov 
Imaging Technique (ACIT)'' . Although considerable effort has been applied 
to the 
development of
alternative techniques, they are not yet competitive and will not be considered here.
 
In this description of the Atmospheric Cherenkov Imaging Technique 
there will be four 
sections: a historical review of the ACIT with emphasis on the early days 
in which the technique was established, a brief outline of the general principles underlying
atmospheric Cherenkov telescopes (ACT), a description, albeit incomplete, of the 
ACIT as currently used and the present generation of instruments, and a 
summary, which will rapidly become dated, of the current observational status of the field.
More complete accounts can be found elsewhere \cite{weekes1988},\cite{Aharonian:97},
\cite{fegan1997},\cite{ong},\cite{weekes2003}.

\section{Early History of the Atmospheric Cherenkov Technique}

\subsection{Discovery of the Phenomenon}

\begin{figure}
\begin{center}
\includegraphics[width=.4\textwidth]{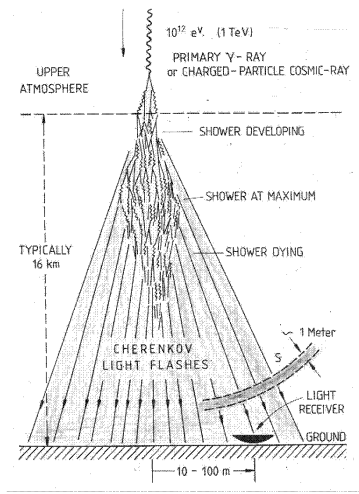}
\end{center}
\caption[]{Cartoon of the atmospheric Cherenkov shower phenomenon, as drawn by
J.V.Jelley in 1993.}
\label{jvj}
\end{figure}

In the Ph.D. dissertations of students studying the atmospheric Cherenkov phenomenon 
the first reference should be to the 1948 note by the British Nobel Laureate, P.M.S. 
Blackett in the Royal Society report on the study of night-sky light and aurora
\cite{blackett}; in that note 
he points out that perhaps 0.01\% of the night-sky light must come from 
Cherenkov light emitted by cosmic rays and their secondary components as they 
traverse the atmosphere. In practice few students actually have read the note and
indeed little attention was paid to this prediction (since it seemed unobservable) 
at the time. Fortunately five years later when Blackett was visiting the Harwell Air Shower
array he brought his prediction to the attention of two Atomic Energy Research
Establishment physicists, Bill Galbraith and John Jelley. After the visit the idea 
occurred to them that, while the net flux of Cherenkov light would be impossible to 
measure, it might be possible to detect a short light pulse from a cosmic ray air shower 
which involved some millions of charged particles (Figure~\ref{jvj}).

Within a week Galbraith and Jelley had assembled the items necessary to test their
hypothesis. A 5 cm diameter photomultiplier tube (PMT) 
was mounted in the focal plane of a 25 
cm parabolic mirror (all housed in a standard-issue Harwell garbage can) and 
coupled to an amplifier with a state-of-the art 5 MHz amplifier whose output was 
displayed on an oscilloscope. They observed oscilloscope triggers 
from light  pulses that exceeded the average 
noise level of the night-sky background every two minutes. They noted that the 
pulses disappeared when the garbage can lid was put in place and a padding lamp 
was adjusted to give the same current in the PMT as was observed from the night-sky
\cite{gj1952}. 
Jelley noted that if the rate had been any lower than that observed they would 
probably have given up and gone home! \cite{jvj1993}. It is not often that a new
phenomenon can be discovered with such simple equipment and in such a short
time, but it may also be true that it is not often that one finds experimental
physicists of this quality!

\subsection{The Power of the Technique}

\begin{figure}
\begin{center}
\includegraphics[width=.3\textwidth]{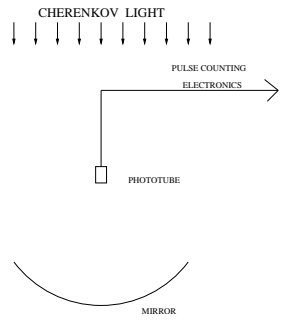}
\end{center}
\caption[]{The essential elements of an Atmospheric Cherenkov Detector}
\label{detector}
\end{figure}

With the Harwell air shower array (one of the largest 
such arrays then in existence) in close
proximity, it was easy to show that the light pulses 
were indeed associated with air showers.
In the years that followed, Galbraith and Jelley  made a series of 
experiments in which they determined the basic parameters of the Cherenkov
radiation from air showers. The account of these elegant experiments is a must-read
for all newcomers to the field \cite{gj1953},\cite{jg1953}. 
The basic detector elements are extremely simple (Figure~\ref{detector}).
It was realized at an early stage that
the phenomenon offered the possibility of detecting point sources of cosmic
ray air showers with high efficiency. Since charged primaries are rendered 
isotropic by the intervening interstellar magnetic fields, in practice this meant
the detection of point sources of neutral quanta, i.e., gamma-ray photons or
perhaps neutrons. 
The lateral spread of the Cherenkov light from the shower 
as it strikes the ground is $\approx$ 100-200 m 
so that even a simple light receiver of modest dimensions has an effective collection 
area of some tens of thousands of square meters. The fact that the light pulse 
preserves much of the original direction of the primary particle and that the
intensity of light is proportional to the total number of secondary particles, 
and hence to the energy of the primary, makes the detection technique potentially
very powerful.

\begin{figure}
\begin{center}
\includegraphics[width=.4\textwidth]{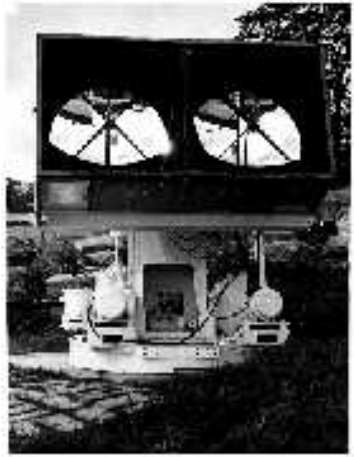}
\end{center}
\caption[]{The British-Irish telescope at Glencullen, Ireland c. 1964; the telescope
consisted of two 90 cm searchlight mirrors on a Bofors gun mounting.}
\label{glencullen}
\end{figure}

\begin{figure}
\begin{center}
\includegraphics[width=.5\textwidth]{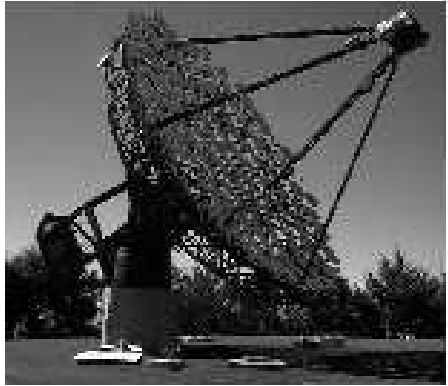}
\end{center}
\caption[]{The Whipple Observatory 10 m gamma-ray telescope was built in 1968; 
it is still in operation. It is composed of 250 glass facets, each of focal 
length 7.3 m.}
\label{10mtelescope}
\end{figure}

The prediction by Cocconi \cite{cocconi1959} of a strong flux of 
TeV gamma rays from the Crab 
Nebula precipitated an experiment by the Lebedev Research Institute in the Crimea
in 1960-64 \cite{chudakov1965}.  Supernova 
Remnants and Radio Galaxies had recently been identified as sources containing 
synchrotron-emitting electrons which suggested that they might be gamma-ray sources.
A selection of these (including the Crab Nebula) were examined with a simple ACT
 system which did not 
attempt to discriminate between air showers initiated by gamma rays and those 
initiated by hadrons. No sources were found but the basic methodology 
involved in a search for point source anisotropies in the cosmic ray air shower
distribution was defined. The technique was refined John Jelley and Neil Porter 
in a British-Irish experiment \cite{long1964} 
 in which the candidate source list was expanded to include the recently discovered 
quasars and magnetic variable stars (with null results). All these early 
experiments used ex-World War II searchlight mirrors (Figure~\ref{glencullen}. 
The first purpose-built 
optical reflector for gamma-ray astronomy was the Smithsonian's 10 m reflector
on Mount Hopkins in southern Arizona (Figure~\ref{10mtelescope}). This telescope, built
by Giovanni Fazio, was the first purpose-built gamma-ray telescope; it is still
in use after 37 years. This again was a first generation device in
which the assumption was made that there was no easily measured differences in the
light pulses from gamma-ray and hadronic primaries. The motivation for this large
increase in mirror area (and decrease in energy threshold) was a refined prediction
of a detectable flux of gamma rays from the Crab Nebula based on a 
Compton-synchrotron model \cite{gould1964}.

Although these first generation detection systems were extremely simple and 
exploited the ease with which gamma rays could be {\it detected}, they did not
provide the means of {\it identifying} gamma rays among the much more numerous
cosmic ray background 
Hence, until 1989 when the Crab Nebula was detected
\cite{weekes1989}, there was no credible detection of a 
gamma-ray flux from any cosmic source.

\subsection{Basic Principles}

Some feel for the quantities involved in Cherenkov light emission from air showers
in the energy range of interest can be seen from Table~\ref{cascade} based on 
Monte Carlo simulations by A.M. Hillas. 
Note that the various quantities
$N_{max}$, the number of particles in the shower at the shower maximum and
$N$ and $\rho$, the number of particles and optical photons at sea level and at 
mountain altitude (2.3 km) scale with primary energy.
 
\begin{center}
\begin{table}
\caption{Shower Parameters as a Function of Energy \cite{weekes2003}
\label{cascade}
}
\begin{tabular}{rrrrrrrr}
\hline
Energy, E$_\gamma$ & X$_{max}$ & h$_{max}$ &   $N_{max}$ & $N_{sl}$
& $N_{mt}$ & $\rho_{sl}$ & $\rho_{mt}$\\
     & g cm$^{-2}$ &  km&  &   &  & ph-m$^{-2}$& ph-m$^{-2}$ \\
\hline
100 GeV & 261 &  10.3   & 130  & 0.04 & 1.4 &  4  & 8    \\
1 TeV   & 346 & 8.4  &  1,140  & 3  & 60 &  74  & 130 \\
10 TeV  & 431 & 6.8 & 10,000 & 130  & 1,700 & 11,000 & 1,700       \\
100 TeV & 517 & 5.5 & 93,000 & 4,500 & 36,000 & 16,000 & 19,000\\
\hline
\end{tabular}
\end{table}
\end{center}

The light signal (in photoelectrons) detected is
given by:\\
S = $\int^{\lambda_1}_{\lambda_2}$ k E($\lambda$)  T($\lambda$) $\eta(\lambda$) A
d$\lambda$ \\
where C($\lambda$) is the Cherenkov photon flux within the
wavelength sensitivity
bounds of the PMT, $\lambda_1$ and $\lambda_2$,
E($\lambda$) is the shower Cherenkov emission spectrum
(proportional to 1/$\lambda^2$), T($\lambda$) is the atmospheric
transmission  and k is a constant which
depends on the shower, and the geometry.

 The signal must be detected above the
fluctuations in the night-sky background during 
the integration time of the pulse counting system, $\tau$.

The sky noise B is given by:\\
B = $\int^{\lambda_1}_{\lambda_2}$ B($\lambda$) $\eta$($\lambda$)
$\tau$ A $\Omega$ d$\lambda$.\\
Hence the signal-to-noise ratio is essentially\\
S/N = S/B$^{0.5}$  = $\int^{\lambda_1}_{\lambda_2}$ C($\lambda$)
[$\eta$
($\lambda$) A /$\Omega$ B($\lambda$) $\tau$]$^{1/2}$ d$\lambda$.

The smallest detectable light pulse is inversely proportional to
S/N; the minimum detectable gamma ray then has an energy threshold,
E$_T$ given by\\
E$_T$ $\propto$ 1/C($\lambda$) [B($\lambda$) $\Omega$
$\tau$/{$\eta$($\lambda$) A]$^{1/2}$

If S = the number of gamma rays detected from a given source in a
time, t,
and  A$_\gamma$ is the collection area for gamma-ray detection,
then
S = F$_\gamma$(E) A$_\gamma$ t.
The telescope will register a background, B, given
by:\\
B = F$_{cr}$ A$_{cr}$(E) $\Omega$ t,
where A$_{cr}$(E) is the collection area for the detection of
cosmic rays of energy E.
The cosmic ray background has a
power law spectrum:\\
F$_{cr}$($>$E) $\propto$ E$^{-1.7}$ and if we
assume the gamma-ray source  has the form:\
F$_\gamma$($>$E$_\gamma$) $\propto$ E$_\gamma$$^{-a_\gamma}$.

Then the standard deviation,
$\sigma$ $\propto$ S/B$^{1/2}$    $\propto$ E$^{1.7/2-a_\gamma}$
[A$_\gamma$/A$_{cr}$]$^{1/2}$t$^{1/2}$

The minimum number of standard deviations, $\sigma$, for a reliable
source detection is generally taken as 5 \cite{weekes2003}.

\section{Early Development of the ACIT}

\subsection{Discrimination Methods}

At an early stage it was realized that while the atmospheric Cherenkov technique
 provided a very easy way of {\it detecting} 
gamma rays with simple light detectors, it did not 
readily provide a method of discriminating the light pulse from gamma-ray air showers 
from the background of light pulses from the much more numerous cosmic ray showers; 
thus the {\it flux sensitivity} was severely limited.  
Although these are isotropic, there is typically a ratio of  1,000-10,000 of cosmic 
rays to gamma rays recorded by the simple light detectors that were 
available in the two decades following the Harwell experiments.
Once it was apparent that the early, very optimistic, predictions of the strength 
of the most obvious potential TeV sources were not to be realized, then attention turned to 
methods of improving the flux sensitivity of the technique. Although superficially very 
similar, Monte Carlo simulations of shower development and Cherenkov light emission 
suggested some differences that might be exploited to preferentially select gamma rays.
 
These differences are listed below and illustrated in the cartoons
 in Figure~\ref{parameters}:
\begin{itemize}
\item Lateral Spread at ground level: the light pool from gamma-ray showers is more 
uniform than that from cosmic ray showers. This feature is difficult to exploit since it 
requires numerous light detectors spread over relatively large areas; it has recently been  
used by the group at the Tata Institute at their Pachmari site \cite{pachmari}
\item Angular Spread: the image of the light superimposed on the night-sky 
background has a more regular distribution from gamma-ray showers and is smaller and more 
uniform. This feature was recognized by Jelley and Porter 
\cite{jelleyporter1964} but not really exploited 
until some decades later. This was to prove the most powerful discriminant and to lead to 
the first successful credible detection of a TeV gamma-ray source \cite{weekes1989}.
\item Time Structure: because the cosmic ray component contains penetrating particles 
(mostly muons) that survive to detector level, the duration of the light pulse can be 
longer. Many early versions of the ACT, particularly the Haleakala experiment 
\cite{haleakala}, 
attempted to exploit this feature but it was not to prove very effective,
\item  Spectral Content: the penetrating component of cosmic ray showers is close to 
the light detector and its overall Cherenkov light at the detector is less attenuated 
in the ultraviolet; this feature was used as a discriminant in the early 
Whipple and Narrabri 
experiments of Grindlay and his collaborators \cite{grindlay1975}
 and in the Crimean experiments 
\cite{crimea}. It is mostly effective when combined with other discriminants.
\end{itemize} 

\begin{figure}
\begin{center}
\includegraphics[width=.4\textwidth]{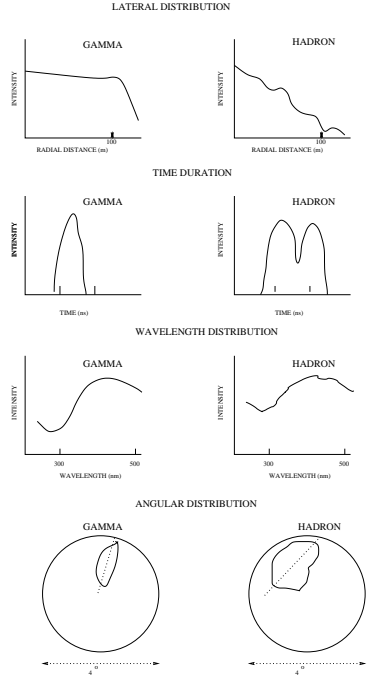}
\end{center}
\caption[]{Cartoon depiction of parameters used to discriminate Cherenkov light from
gamma-ray and hadron air showers}
\label{parameters}
\end{figure}

The Cherenkov light image has a finite angular size which can, in principle,
be used to refine the arrival directing, and perhaps even to distinguish it from
the images of background cosmic rays \cite{jelley},\cite{jelley1965}. 
However when a simple telescope with a single light  detector (pixel) is used as a 
gamma-rays detector, this information is lost and the angular resolution
is no better than the field of view of the telescope. 
Because the Cherenkov light images
are faint and fast, it is not technically straight-forward to record them. 
Boley and his collaborators \cite{boley} had used an array of photomultipliers at Kitt Peak
to study the longitudinal development of large air showers but these were from very
energetic primaries. A
pioneering effort by Hill and Porter 
\cite{hillporter1960}, using a image intensifier system 
from a particle experiment, resulted in the first recorded images of Cherenkov
light from air showers. However, because of the finite size of the photocathode,
it was only possible to couple it to a relatively small mirror which meant
that only cosmic ray primaries above 100 TeV could be detected. The 
potential advantages of this approach as a means of separating out the gamma-ray 
component were recognized \cite{jelleyporter1964}, but since the 
technique was limited to energies where the attenuation of the gamma-ray flux by
photon-photon pair production in intragalactic space was appreciable, this 
approach was not pursued. 

A more practical approach was that pursued by Grindlay and his colleagues 
\cite{grindlay1975} in 
which multiple light detectors separated by distances $\approx$ 100 m
 were used to detect the shower maximum 
associated with gamma-ray showers and the penetrating, mostly muon, component
from hadron showers. The latter was used as a veto to preferentially select events
that were initiated by gamma rays.   
This ''Double Beam'' technique was 
potentially powerful but was difficult to implement with the resources available at the 
time; it received new life when the Narrabri Stellar Interferometer (in Australia) became 
available. With two large reflectors of 9 m aperture on a circular rail system, the system 
was ideally suited for this technique. Although some detections were reported (the Crab 
pulsar, the Vela pulsar and Centaurus A) \cite{grindlay}, they have not been confirmed 
by later, more sensitive, observations and this technique was not pursued any further.

Activity in ground-based gamma-ray astronomy was at a low ebb in the seventies. 
Observations with the Whipple 10 m reflector had moved the energy threshold of the 
technique close to 100 GeV but this had only produced upper limits on the predicted sources. 
Smaller telescopes produced tentative detections of several binaries and pulsars but these 
were always on the edge of statistical credibility and were not subsequently verified (for 
reviews of this controversial epoch of TeV gamma-ray astronomy, 
see \cite{chadwick}, \cite{weekes1991}).

\subsection{The Power of the Atmospheric Cherenkov Imaging Technique}

The concept of using electronic cameras consisting of matrices 
 of phototubes in the focal plane of large reflectors to record 
the images of the Cherenkov light from small air showers was first suggested in a paper 
at a workshop in Frascati Italy  \cite{weekesturver1977}. Entitled     
''Gamma-Ray Astronomy from 10-100 GeV: a New Approach'' the emphasis was on lowering
the energy threshold through the use of two large reflectors separated by 100 m, each 
equipped with arrays of phototubes in their focal plane. The motivation to go to lower
energies came from the prediction from Monte Carlo simulations that the ratio of Cherenkov
light from gamma-ray showers to cosmic ray showers of the same energy drops off 
dramatically below 100 GeV. In this paper the physical explanation of this falloff 
was stated: ''In a proton shower
most of the Cherenkov light comes from the secondary electromagnetic cascades.
Energy comes into these cascades via the production of pions by the primary 
and the subsequent nucleon cascade. Two thirds of the energy (approximately) 
goes to charged pions; they can decay to muons or undergo a collision.The 
latter process is a more efficient method of producing Cherenkov light; 
since the lifetime against decay is greater a higher energies, the 
chance of collisions is greater. At lower energies therefore, proportionally 
more energy comes off in muons 
whose energy may be below the Cherenkov threshold and hence the low energy 
showers are deficient in Cherenkov light''. 

The idea of using an array of phototubes with limited resolution to image the 
Cherenkov light rather than the high resolution offered by image intensifiers was 
motivated by the experience of the author using CCD detectors in optical astronomy
where the resolution achieved is significantly greater than the scale of the 
pixels.
In the paper there was little emphasis on discrimination of the primaries based on 
the shapes of the images although it was claimed that there would be a significant 
improvement in angular resolution (to 0.25$^\circ$). The use of two reflectors in
coincidence was advocated to reduce the predicted muon background.

In this paper \cite{weekesturver1977} the basic concept of the Cherenkov light
 imaging telescope was described; it consisted of an array of PMTs in
the focal plane of a large reflector. The use of an array of at least two
such cameras was advocated. This has been the model for all future 
telescopes using the ACIT. In general,
in recording the Cherenkov light image from an air shower, the gamma-ray astronomer
tries to characterize its nature (gamma-ray or hadron), determines its arrival direction, 
and gets some estimate of the primary that initiated the air shower. The geometry
of the shower images is demonstrated in Figure~\ref{geometry}. The factors that cause 
the observed shape and size of the image are many: the nature of the primary particle, its 
energy and trajectory, the physical processes in the particle cascade (principally pair
production and bremsstrahlung in electromagnetic cascades with the addition of pion 
production in hadron initiated cascades), Coulomb scattering of shower electrons, 
the effect of geomagnetic deflections of the 
shower particles, the distance of the point of impact of the shower core from the
optic axis, the Cherenkov angle of emission, and the effect of atmospheric absorption
\cite{fegan1997}.
In addition the properties of the imaging system must be completely understood: the
reflectivity of the mirrors, the quantum efficiency of the light detectors as a function
of wavelength, the time response of the system and the distortions introduced by the
system's optics, cables, electronics and data readout.
 
\begin{figure}
\begin{center}
\includegraphics[width=.3\textwidth]{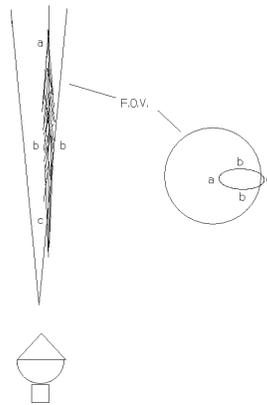}
\end{center}
\caption[]{The geometry of atmospheric Cherenkov imaging. On the left is a 
cross-section of the shower intersection with the field of view of the
light detector whose inverted image plane is seen on the right.}
\label{geometry}
\end{figure}

Fortunately all of these factors are amenable to calculation or measurement. The physics 
of the various processes involved in the shower development are well known and Monte
Carlo methods can be used to estimate the expected values from particular primaries.
However since fluctuations play a major role in such development the expected values
cover a range of possibilities and identification must always be a statistical process.
It is relatively easy to predict the properties of the gamma-ray initiated showers; it is
more difficult to predict the expected properties of the background which is mainly from
charged cosmic rays. While every attempt is made to estimate both signal and background, 
it is usually found that the background contains some unpleasant surprises; hence while the 
gamma-ray detection rate can be reliably predicted, the efficiency of the identification
of the gamma-rays from the more numerous background requires the system to be actually
operated in observations of a known source. Since the background is numerous and
constant, its properties can be readily modeled from empirical databases.
There is an irreducible background from hadron showers which develop like 
electromagnetic cascades (most of the energy goes into a $\pi^o$ in the first interaction)
and from the electromagnetic cascades produced by cosmic electrons (whose fluxes in the
range of interest are 0.1 - 0.01\% of the hadron flux).

\subsection{The First Source}

When the imaging systems first went into operation it was not immediately obvious how
the images should be characterized and discriminated from the background. There were 
no credible sources and Monte Carlo calculations were still being developed
and were untested.
The first such calculations available to the Whipple Collaboration indicated that
fluctuations might effectively rule out any discrimination and did not encourage 
the development of sophisticated analysis techniques. The first Whipple camera had 37 
pixels, each of 0.25$^\circ$ diameter \cite{cawley}. A relatively simple image parameter,
{\it Frac2}, defined as the ratio of the signal in the two brightest pixels to the
total light in the image, was developed empirically and led to the first indication
of a signal from the Crab Nebula \cite{cawley1985}, \cite{gibbs}. This simple parameter
picked out the compact images expected from electromagnetic cascades but did not
provide any information on the arrival direction (other than that it was within
the field of view of the detector). However the application
of the same selection method on putative signals from the then popular sources, Cygnus
X-3 and Hercules X-1, did not improve the detection credibility and initially cast doubt 
on the effectiveness of {\it Frac2} as a gamma-ray identifier.   

Since the images were roughly elliptical in shape, an attempt was made to quantify the
images in terms of their second and third moments \cite{mckeown1983}. However this was not
applied to gamma-ray identification until Hillas  undertook a new series of Monte
Carlo calculations \cite{hillas1985}. 
These calculations predicted that gamma-rays images could be
distinguished from the background of isotropic hadronic images based on two criteria:
the physics of the shower development was different leading to smaller and better defined
ellipses for gamma rays and that the geometry of image formation 
led to all images coming from a point source on axis having their major axes 
intersecting the center of the field of view.
Fortunately the first property aids the definition of the second and provides potentially
very good angular resolution. Hillas \cite{hillas1985} defined  a series of parameters
which included the second moments ({\it Width} and {\it Length}), the parameter
{\it Dist} which measures the distance of the centroid of the image from the optic axis,
and {\it Azwidth} which measures the projected width of the image on the line joining the
centroid to the center of the field of view. Later {\it Alpha}, the angle between this
line and the major axis was added as was {\it Asymmetry}, the third moment. {\it Azwidth}
was particularly simple; it is easy to use and proved to be very effective as it combined 
discrimination based on image size (physics) and arrival direction (geometry) and led 
to the first definite detection of a point source of TeV gamma-rays. In general 
multiple parameter selections were made. The 
parameters were first defined in Monte Carlo calculations but once the standard candle
of the Crab Nebula was established \cite{weekes1989}, optimization was made on the strong 
and steady Crab
signal to preferentially select gamma rays. This optimization led to an analysis package
called Supercuts \cite{punch}, which proved to be extraordinarily robust, and in various
forms, was the basis of the data analysis used by the Whipple Collaboration to detect 
the first AGN 
\cite{punch1992},\cite{quinn1996}, \cite{holder},\cite{catanese},\cite{horan2003}. 
Other groups have defined different
parameters and analysis schemes but the basic methodology is the same. 

\section{ACT Observatories}

\subsection{Third Generation Observatories}

By 1996 the ACIT was judged to have been very successful and a number of groups
made plans for a third generation of the ACTs. The limitation of a single telescope
was easily seen from the results obtained using the Whipple telescope and camera 
\cite{kildea05}. At low trigger thresholds it was impossible to distinguish low
energy gamma-ray events from the much more numerous background of partial muon 
rings (arcs). Despite intense efforts with sophisticated analysis methods it was 
clear that the discrimination threshold was a factor of 2-3 above the trigger 
threshold. Hence although the fundamental threshold was $\approx$ 200 GeV, the effective 
threshold was $\approx$ 400 GeV. Since the muon Cherenkov emission is essentially a 
local phenomenon, this background is easily eliminated by demanding a
coincidence with a second telescope separated from the first by a minimum
distance of 50 m \cite{weekesturver1977}. In fact the HEGRA experiment had 
already demonstrated \cite{hegra} the power of an array of small imaging telescopes
to improve the angular and energy resolution of the ACIT; at the threshold 
energies of these telescopes the muon background was not a problem. 

Thus it was apparent that the next generation of the ACIT 
 would involve arrays of reflectors
with apertures in excess of 10 m, with better optics, with more sophisticated 
cameras, and with data acquisition systems capable of handling high rates. Such 
systems required an investment that was almost an order of magnitude greater 
than the previous generation of detectors (but the flux sensitivity was to
improve by a similar factor (Figure~\ref{sensitivity})). Of necessity the number
of people involved in each experiment would be so large ($\approx$ 100)
that the new collaborations
would be more in line with the numbers of scientists found in particle physics
experiments than in typical large astronomical projects.

\begin{figure}
\begin{center}
\includegraphics[width=.7\textwidth]{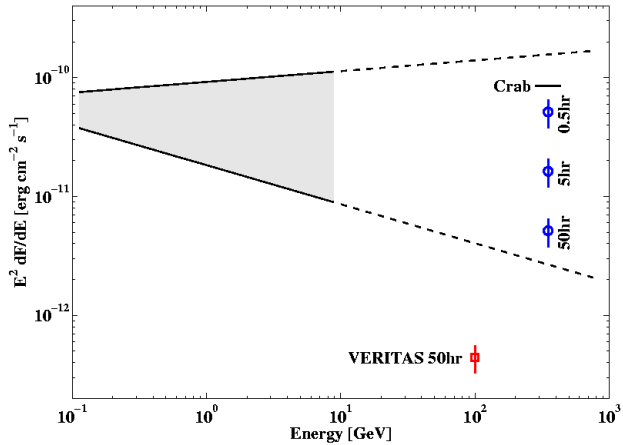}
\end{center}
\caption[]{Integral flux sensitivity for various integration times for the Whipple
telescope (unlabeled points) and for the projected VERITAS array \cite{sfegan}.
The dotted lines are the extrapolated energy spectra from sources detected by
EGRET.}
\label{sensitivity}
\end{figure}

\subsection{The Power of ACT Arrays}

ACTs arrays can be discussed under the headings of improvements offered in energy threshold,
energy resolution, angular resolution and background discrimination. A good discussion 
can be found in \cite{Aharonian:97}. A typical array provides multiple images of
a single event as seen in Figure~\ref{cartoon}.

\begin{figure}
\begin{center}
\includegraphics[width=.7\textwidth]{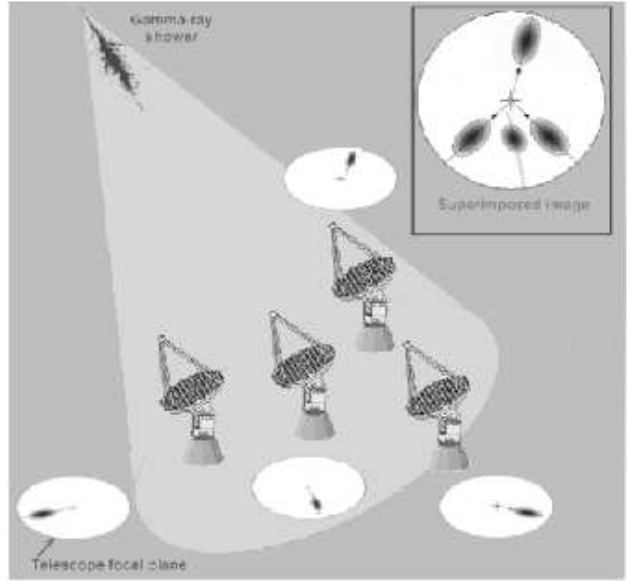}
\end{center}
\caption[]{Cartoon showing response of array of four detectors to
air shower  whose axis is parallel to the optical axes of the
telescopes and some 30 m displaced from the center of array.
(Figure courtesy of P.Cogan)}
\label{cartoon}
\end{figure}

\noindent
{\bf Energy Threshold:}
The basic quantities involved in determining the energy threshold of an ACT 
are given above in Section 2.3 and are fairly obvious: the mirror area should be as 
large as possible and the light detectors should have the highest possible
quantum efficiency. To the first approximation (as demonstrated in \cite{chudakov1965})
it does not critically depend on how the mirror area is distributed, i.e., a cluster
of small telescopes in close proximity operated in coincidence is the same as if their
signals are added and is approximately the same as that of a single large telescope
of the same total mirror area. Practical considerations tend to dominate: coincidence
systems are more stable, the cost of telescopes scales as the (aperture)$^{2.5}$, the
relative cost of multiple cameras on multiple small telescopes versus the 
cost of a single camera on a large telescope, etc. However the simplest way to get
the lowest energy threshold is to go for a single large telescope (although this
may introduce other problems).

\noindent
{\bf Angular Resolution:}
Angular resolution is important not only for reducing the background and identifying
a potential source but also for mapping the distribution of gamma rays in the 
source.
Stereoscopic imaging, the simplest form of "array" imaging, offers the immediate
advantage of improving the angular resolution. This principle was established
with the use of just two telescopes with a separation of $\approx$ 100 m, i.e., with the
two telescopes within the light pool of the Cherenkov light pool, $\approx$
a circle of diameter 200 m. The greater the separation, the better the
angular resolution but increasing the separation beyond 100 m begins to reduce
the effective gamma-ray collection area. A simple array of imaging 
ACTs can provide a
source location of $\approx$ 0.05$^\circ$ for a relatively strong source with angular 
resolution of $\approx$ 0.1$^\circ$ for individual events. This is a factor of two 
improvement over that for a single telescope.  
An angular resolution of an arc-min or better appears feasible.

\noindent
{\bf Background Discrimination:}
Multiple views of the same air shower from different angles obviously improves
the signal-to-noise ratio when the images are combined. However in reducing the
background of hadronic events the gain is not as large as might appear at 
first glance. Hadronic showers which develop like typical showers are 
easily identified and rejected, even in a single telescope. More subtle are
the hadronic events which develop like an electromagnetic cascade (an
early interaction channels much of the energy into an electron or
gamma ray). Such events cannot be identified not matter how many views are 
provided on the cascade development. Similarly the cascades initiated by 
cosmic electrons are an irreducible background. However the array approach 
does completely remove the background from single local muons and the 
improved angular resolution narrows the acceptable arrival directions. 

\noindent
{\bf Energy Resolution;}
The Cherenkov light emitted from the electromagnetic cascade is to a first
approximation proportional to the energy of the initiating gamma ray 
(Table~\ref{cascade}). However
with a single ACT there is no precise information as to the impact parameter
of the shower axis at ground level. Since the intensity of the Cherenkov light 
is a function of distance from the shower axis, the lack of information on this
parameter is the limiting factor in determining the energy of the gamma ray. The
energy resolution of a single imaging ACT is $\approx$ 30-40\%.
With an array the impact parameter can be determined to $\approx$ 10 m and the energy
resolution can be reduced to 10\%.

\subsection{The Third Generation Arrays}

This third generation of ACTs has seen the formation of four
large collaborations formed to build  arrays of large telescopes: 
a largely German-Spanish 
collaboration that is building two 17 m telescopes on La Palma in 
the Canary islands (MAGIC) \cite{magic} (Figure~\ref{magic}): an
Irish-British-Canadian-USA collaboration that is building an array of four
12 m telescopes in Arizona (VERITAS) \cite{veritas}; an Australian-Japanese
collaboration that has built four 10 m telescopes in Australia
(CANGAROO-III) \cite{cangarooIII}; a largely European collaboration that has built
an array of four 12 m telescopes in Namibia (HESS) \cite{hess} 
and plans to add a fifth 
telescope of 28 m aperture at the center of the array
(Figure~\ref{hess}). The fact that two of the arrays are in each hemisphere
is somewhat fortuitous but ensures that there will be good coverage of the
entire sky and that all observations can be independently verified. 
The principal properties of the four arrays are summarized in Table~\ref{arrays}. 

The sensitivity of these new arrays is probably not dissimilar but only HESS
has demonstrated what it can achieve in the actual detection of known sources.
With the second generation of ACTs (Whipple, HEGRA), it was possible to detect
a source that was 5\% of the Crab Nebula in 100 hours of observation. With 
HESS this is reduced to one hour and in principle in 100 hours it should be 
possible to detect a source as weak as 0.5\% of the Crab. HESS has also 
demonstrated an energy resolution of 10\% and an angular resolution of
an arc-min.
 
\begin{table}
\begin{center}
\caption{Next Generation ACT Arrays}
\label{arrays}
\begin{tabular}{lllllll}
Experiment & Location & Elevation & Telescopes & Aperture &
Pixels & Energy \\
          &         & km & & m &/camera & GeV \\
\hline
CANGAROO-III& Woomera, Australia & 0.2 & 4 & 10 & 577 & 50? \\
HESS & Gamsberg, Namibia & 1.8 & 4 & 12 & 960 & 50 \\
MAGIC & La Palma, Spain & 2.3 & 2 & 12 & 577 & 20? \\
VERITAS &  Arizona, USA & 1.8 & 7 & 12 & 499 & 50 \\
\hline
\end{tabular}
\end{center}
\end{table}

\begin{figure}
\begin{center}
\includegraphics[width=.6\textwidth]{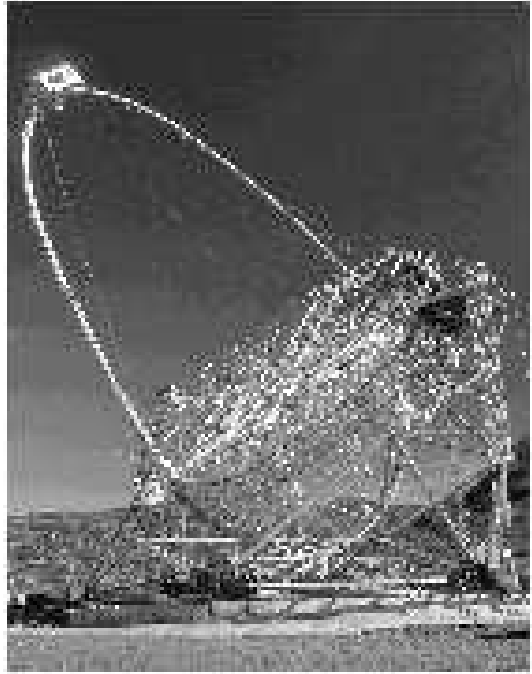}
\end{center}
\caption[]{The 17m aperture telescope, MAGIC, which was completed in 2003
and is now in operation on La Palma in the Canary Islands, Spain.}
\label{magic}
\end{figure}

\begin{figure}
\begin{center}
\includegraphics[width=.8\textwidth]{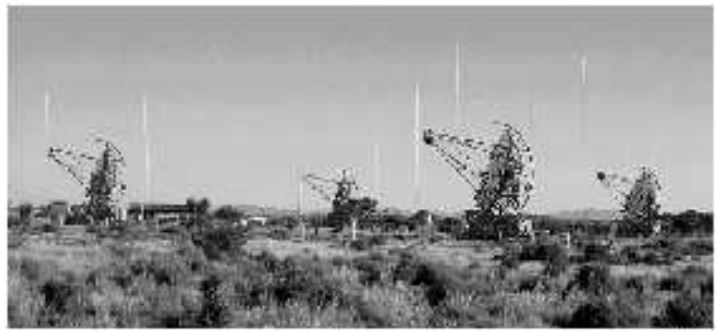}
\end{center}
\caption[]{The HESS array of four 12m aperture telescopes; it has been in full
operation in Namibia since 2004.}
\label{hess}
\end{figure}

\subsection{Hardware Considerations}

{\bf Location:}
Although it is generally accepted that ACTs gain sensitivity by going 
to higher elevations, practical considerations have to date
limited such observatories to quite moderate elevations by astronomical
standards. The CANGAROO-III observatory is near sea level and the other 
three observatories are at conventional optical astronomical elevations 
($<$ 2.5 km).

{\bf Number and Configuration:}
The minimum number of telescopes is more than 
two. Three is optimum but four gives some redundancy and is usually the
preferred number.
Monte Carlo simulations indicate that the telescope arrays are somewhat 
insensitive to the precise configuration of the telescopes and to the 
exact separation. 
The HESS telescopes have a square configuration of side 100 m. The 
CANGAROO-III telescopes are in a diamond-shaped configuration with
characteristic spacing of 120 m. Three of the VERITAS telescopes form a 
equilateral triangle with a fourth at the center; 
the distance from it to each of the corners is 80 m.    

{\bf Light Detectors:}
It is remarkable that the same light detectors are in use in all of this 
generation of experiments as were used in the initial experiments 
of Jelley and Galbraith fifty years ago. The remarkably robust PMT
tube is fast, has a high gain and is reasonably efficient; however it requires
operation at high voltages, has quantum efficiencies less than 25\% and 
is easily damaged by exposure to bright light. For 
many years it has appeared that it is about to be replaced by a new technology
device with higher quantum efficiency. However as of this date 
no such device has yet been used in any application
to detect Cherenkov light pulses from air showers. Hence the practical 
application of such devices to the rather demanding cameras on ACTs still 
seems some way off.

{\bf Optics:}
Mirror area is a critical factor in ACTs. It is not practical to use a 
single large mirror because the cost of producing and supporting it is so
large (although the CANGAROO group used the 8m Subaru optical
telescope in Hawaii for a short period). Mirror area therefore is usually achieved
by the use of multiple facets, which are relatively light, are economical 
to produce, can be recoated easily and can be close packed to mimic a 
single mirror surface. However the facets introduce aberrations and require
careful alignment. The facets can be circular (HESS, CANGAROO), square (MAGIC)
or hexagonal (VERITAS); the circular shape is inefficient but is cheaper to
produce. Glass is still the generally preferred material (HESS, VERITAS) with 
the optical figure formed by slumping, then grinding and polishing; MAGIC uses
diamond-machined aluminum mirrors and CANGAROO-III uses composite plastic
mirrors which, 
although lighter, do not have the optical quality of glass mirrors. The 
aluminum coating on the latter must be overcoated with quartz (HESS) or
anodized (VERITAS).

{\bf Positioners:}
ACTs have not used equatorial mounts. The MAGIC and HESS experiments have 
utilized positioners with the
alt-azimuth design used in some large radio telescopes and solar energy devices;
the elevation motion is a large circular gear while the azimuth motion is rotation
around a large track that is the diameter of the aperture. The CANGAROO-III 
and VERITAS telescopes use the conventional alt-azimuth mounting used in the Whipple
10 m reflector and in many radio telescopes and communication devices.    

\section{Observation Summary}

VHE gamma-ray astronomy is now a fast moving field and the observational picture 
is changing quickly as the new generation of telescopes comes on-line. The HESS
observatory has been particularly productive and it is expected that it will shortly
be joined by CANGAROO-III, MAGIC and VERITAS. The catalog of current sources
listed in Table~\ref{catalog} is dominated by HESS results; it is a measure of their
success that they have been able to announce a new discovery every month and no end
appears to be in sight. As in any rapidly developing field this catalog will rapidly 
become out of date; it is current as of July, 2005 and is quite different from that
presented at the Heraeus school twelve months earlier.
Entries to the table are based on published results in
refereed journals. The first column gives the catalog name for the source, the 
second the conventional source name (where there it is a known object), the 
third is the source type, where known, the fourth the group responsible for the 
discovery, the fifth the date of discovery, and finally the significant discovery reference. 
A feature of this new catalog is that not only does it contain many new sources 
compared with previous listings \cite{weekes2003}, but it also contains some 
significant omissions. Several sources, including TeV 0047-2518 (NGC 253), 
TeV 0834--4500 (Vela), TeV 1503-4157 (SN1006) and TeV 1710-2229 (PSR 1706-44),
have not been verified by the more sensitive observations by HESS. All four sources
were reported with good statistical significance by the CANGAROO group and it 
is a matter of concern in the VHE gamma-ray community that these sources were
reported and published in refereed journals. It is 
apparent that there were unknown systematic errors in the data taking and/or 
the analyzes were not independently verified within the large CANGAROO
collaboration.
It is unfortunate for the discipline that these important sources, whose 
discovery had been greeted with some excitement, have been red herrings and
have decreased the credibility of other legitimate discoveries. Since many 
of these CANGAROO pseudo-sources were reported to have steep spectra, one
possible explanation for the data was unevenly matched ON and OFF fields
and hence systematic biases in the datasets. 

It should be noted that a few of the sources listed in Table~\ref{catalog}
still do not have the statistical significance and independent verification
that one would like. These include TeV 0219+4248 (3C66a), TeV 1121-6037
(Centaurus X-3), TeV 2203+4217 (BL Lac) and TeV 2323+5849 (Cassiopeia A).

\begin{table}
\caption{Source Catalog c.2005}
\label{catalog}
\begin{center}
\begin{tabular}{llllll} 
\hline
\\
TeV Catalog   & Source       &   Type     &   Discovery  &  & Reference \\
Name          &              &            &   Date &  Group &   \\
\hline
TeV 0219+4248 & 3C66A        & Blazar     & 1998   & Crimea & \cite{3C66a}  \\
TeV 0535+2200 & Crab Nebula  & SNR        & 1989   & Whipple& \cite{weekes1989} \\
TeV 0852-4622 & Vela Junior  & SNR        & 2005   & HESS &\cite{0852.0-4622}\\
TeV 1121-6037 & Cen X-3      & Binary     & 1998   & Durham & \cite{chadwick98}  \\
TeV 1104+3813 & Mrk\,421     & Blazar     & 1992   & Whipple & \cite{punch1992}\\
TeV 1231+1224 & M87          & Radio Gal. & 2003   & HEGRA& \cite{m87}  \\
TeV 1259-63 & PSR1259-63/SS2883 & Binary Pulsar& 2005 & HESS  &\cite{1259-63}\\
TeV 1303-631? & Unidentified &  SNR?        & 2005 & HESS &\cite{1303-631}\\
TeV 1429+4240 & H1426+428    & Blazar     & 2002   & Whipple & \cite{horan01}   \\
TeV 1514-5915 & MSH15-52     & PWN        & 2005   & HESS    & \cite{msh15-52}\\
TeV 1614-5150 & Unidentified & ?          & 2005   & HESS & \cite{survey} \\
TeV 1616-5053 & PSR1617-5055? & Pulsar    & 2005   & HESS & \cite{survey} \\
TeV 1640-4631 & G338.-0.0    &  SNR       & 2005   & HESS & \cite{survey} \\
TeV 1654+3946 & Mrk\,501     & Blazar     & 1995   & Whipple & \cite{quinn1996} \\
TeV 1712-3932 & RXJ1713.7-39 & SNR        & 1999   & CANGAROO& \cite{muraishi00} \\
TeV 1745-2900 & Gal. Cen.    & AGN?       & 2005   & HESS & \cite{galcen} \\
TeV 1747-2809 & SNR G0.9+0.1 & SNR        & 2005   & HESS & \cite{survey} \\
TeV 1804-2141 & G8.7-0.1/W30 & SNR        & 2005   & HESS & \cite{survey} \\
TeV 1813-1750 & Unidentified & ?          & 2005   & HESS & \cite{survey} \\
TeV 1825-1345 & PSR J1826-1334? & Pulsar  & 2005   & HESS & \cite{survey} \\
TeV 1826-148 & LS 5039       & Microquasar& 2005   & HESS &\cite{msh15-52}\\
TeV 1834-0845 & G23.3-0.3/W41? & SNR      & 2005   & HESS & \cite{survey} \\ 
TeV 1837-0655 & G25.5+0.0?   & SNR        & 2005   & HESS & \cite{survey} \\
TeV 2000+6509 & 1ES1959+650  & Blazar     & 1999   & Tel.Ar. & \cite{nishiyama00} \\
TeV 2005-489  & PKS 2005-489 & Blazar     & 2005   & HESS& \cite{2005-489}\\
TeV 2032+4131 & CygOB2?      & OB assoc. & 2002   & HEGRA&\cite{aharonian02} \\
TeV 2159-3014 & PKS2155-304  & Blazar     & 1999   & Durham & \cite{2155-304} \\
TeV 2203+4217 & BL Lacertae  & Blazar     & 2001   & Crimea& \cite{neshpor01} \\
TeV 2323+5849 & Cas A        & SNR        & 1999   & HEGRA & \cite{casA} \\
TeV 2347+5142 & 1ES2344+514  & Blazar     & 1997   & Whipple & \cite{catanese}   \\
\hline
\end{tabular}
\end{center}
\end{table}

The most complete catalog of sources is that of blazars of the HBL classification;
these are those whose synchrotron spectrum peaks in the X-ray part of the spectrum.
All of these detections are well-established; their principal properties are listed
in Table~\ref{tev1} which is updated from that given in \cite{Horan03}. 
Taken together these sources
form the basis for a new exploration of relativistic particles in AGN jets.
 
\begin{table}
\caption{TeV Blazars}
\begin{center}
\label{tev1}
\begin{tabular}{llcccl} \hline
Source    & Class & Redshift & F$_\gamma$ (mean)          & F$_\gamma$ (Det.)           &  E$_{peak}$ \\
      &     &    &  $>$ 100 MeV               & $>$ E$_{peak}$              & (Det.)      \\
      &      &         & 10$^{-8}$cm$^{-2}$s$^{-1}$ & 10$^{-12}$cm$^{-2}$s$^{-1}$ & GeV         \\
\hline
Mrk\,421 & BL Lac(HBL) & 0.031   & 13.9                       & 15.0  & 500        \\
H1426+428 &  BL Lac(HBL) & 0.129   & U.L.                     & 20.4 & 280        \\
Mrk\,501 & BL Lac(HBL) & 0.034   & U.L.                       & 81   & 300        \\
1ES1959+650  & BL Lac(HBL) & 0.048   & U.L.                       & 29.4 & 600        \\
1ES2005-489 &  BL Lac (HBL)& 0.071& U.L. & 6.9 & 200 \\
PKS2155-304  & BL Lac(HBL) & 0.117    & 13.2                       & 42.0 & 300        \\
1ES2344+514  & BL Lac(HBL) & 0.044   & U.L.                       & 11.0  & 350        \\
\hline
\end{tabular}
\end{center}

\end{table}

\section{Conclusion}

It is clear that TeV sources are ubiquitous and a powerful tool for exploring
the relativistic universe. Despite this rich catalog of sources there is still
not unambiguous evidence for the source of the hadronic cosmic radiation; it 
is possible to explain all the observed TeV gamma rays as coming from
electron progenitors. Hence despite the dramatic advances that the new catalogs
of TeV sources represent, the origin of the cosmic radiation remains a
mystery.

Although the cement is hardly dry in the foundations of the third generation of 
ACTs there is already active discussion of how the fourth generation might be
configured. It appears that in the energy range from 100 GeV to 100 TeV there 
will be no technique, either in space or on the ground, can hope to compete
with the ACIT in terms of flux sensitivity to  point sources in the next 
decade (Figure~\ref{veritas}). It is technically
possible to build VHE observatories that will have flux sensitivities in the
100-1000 GeV range that will exceed those currently achieved by
a factor of ten. These new arrays will be
particularly sensitive to transient emission and hence the focus may be on
cosmological studies. Several concepts involve ACTs that can reach down to 
energies as low as 10 GeV. There do not appear to be any space missions on 
the drawing boards that would offer a major extension to the sensitivity 
of GLAST. At a recent workshop \cite{palaiseau} several interesting concepts
for a new generation ACT were proposed; the ground-based gamma-ray community has not
yet coalesced to a single concept. This consolidation of manpower and resources
will be surely necessary if a project of this
magnitude is to be realized. However all are agreed that a new generation 
observatory is both necessary and feasible.

\begin{figure}
\begin{center}
\includegraphics[width=.7\textwidth]{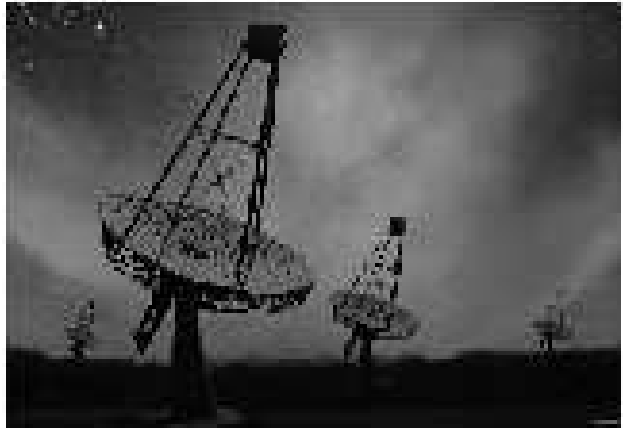}
\end{center}
\caption[]{A photomontage showing the VERITAS observatory; this will have four
telescopes of 12 m aperture and will come on-line in 2005.}
\label{veritas}
\end{figure}

{\bf Acknowledgments:} Over the past 39 years ground-based gamma-ray astronomy
at the Smithsonian's Whipple Observatory has been supported by times by the
Smithsonian Astrophysical Observatory, the Department of Energy, the National
Science Foundation and NASA. Deirdre Horan is thanked for helpful comments on the
manuscript.


\end{document}